\documentclass[preprint,aps,showpacs,preprintnumbers,amsmath,amssymb]{revtex4}

\usepackage{graphicx}
\usepackage{dcolumn}
\usepackage{bm}

\begin{document}

\preprint{APS/123-QED}

\title{Inhomogeneous magnetism in decagonal Al$_{69.8}$Pd$_{12.1}$Mn$_{18.1}$}

\author{D. Rau}
\email{drau@solid.phys.ethz.ch}
\author{J. L. Gavilano}
\author{Sh. Mushkolaj}
\author{C. Beeli}
\author{M. A. Chernikov}
\author{H. R. Ott}
\affiliation{Laboratorium f\"ur Festk\"orperphysik, ETH Z\"urich, 8093 Z\"urich, Switzerland}

\date{\today}

\begin{abstract}
We report the results of measurements of $^{27}$Al and $^{55}$Mn
NMR spectra, the related spin-lattice relaxation rates, and the dc
magnetic susceptibility of the stable decagonal quasicrystal
Al$_{69.8}$Pd$_{12.1}$Mn$_{18.2}$. The temperature-variation of
the magnetic susceptibility $\chi(T)$ reveals that the Mn ions
carry only an average magnetic moment of approximately
$2\mu_{\mathrm{B}}$, and confirms a spin-glass type freezing of
the Mn-moments at $T_f=12$K. The NMR spectra reveal two partially
resolved lines for the $^{27}$Al-nuclei, indicating that there are
two different sets of environments for the Al-sites. The
integrated intensity of the Mn line in the spectra suggests that
about half of the Mn ions carry no magnetic moment. Below 50K, and
upon decreasing the temperature, the $^{27}$Al NMR-linewidth $w$
and the spin-lattice relaxation rate $T_1^{-1}$ both grow with an
increasingly negative slope, as it is often observed and
interpreted as a critical "slowing down" of magnetic moments in
systems approaching a spin-glass transition. Various features,
such as a broad maximum in $T_1^{-1}(T)$ and a slope change in the
$w(\chi)$ plot, both around 120K, suggest a gradual reduction of
the number of Mn moments with decreasing temperature below 120K.
\end{abstract}

\maketitle

\subsection{\label{sec:intro}Introduction}

Since the discovery of quasicrystalline systems by Shechtman et
al. in 1982\cite{QC:ShechtmanBlech}, substantial progress in the
understanding of quasicrystalline structures has been
made\cite{QC:BeeliHoriuchi,QC:SteurerHaibach,QC:BoudardDeBoissieu}.
Quasicrystalline structures are characterised by a high degree of
atomic order, resulting in well defined X-ray and electron
diffraction peaks, but they lack translational periodicity. This
complicates the understanding of the electronic structure and
hence of some important physical properties. The unusual
electronic transport properties are often traced back to the
formation of a pseudogap in the excitation spectrum at the Fermi
level\cite{QC:TangHill,QC:GiannoSologubenko}. The non-periodicity
of the crystal lattice is the reason for special features in the
temperature dependence of the thermal conductivity, reflecting a
general type of Umklapp
scattering\cite{QC:GiannoSologubenko,QC:KaluginChernikov,QC:ChernikovBianchi,QC:KaluginKatz}.
The quasiperiodic atomic arrangement is expected to also influence
the magnetic features of quasicrystals (QCs), and in recent years,
studies of magnetic properties of QCs have been in the focus of a
number of research projects. While the initial studies were mainly
performed on metastable Al-Mn QCs \cite{QC:McHenry,QC:WarrenChen},
a lot of progress has been made since then in preparing stable
high-quality ternary QCs with local magnetic moments in alloy
systems such as Al-Pd-Mn or Al-Mn-Ge. Depending on structure and
stoichiometry, these QCs exhibit ferromagnetism\cite{QC:Yokoyama},
diamagnetism\cite{QC:HippertAudier},
paramagnetism\cite{QC:GavilanoRau, QC:Dolinsek} or spin-glass
phenomena\cite{QC:FujimakiMotoya,QC:Chernikov,QC:Lasjaunias}.
Other intensively investigated families of magnetic QCs include
Al-Cu-Fe and Al-Co-Ni alloys\cite{QC:Yamamoto}, and the more
recently synthesized quasicrystalline compounds containing
rare-earth (RE) ions, such as RE-Mg-Zn with well localized
$4f$-electron
moments\cite{QC:FisherCheon,QC:SatoTakakura,QC:GiannoRE,QC:WesselAnuradha}.

Several studies of the $d$-electron magnetism of Mn in icosahedral
samples of Al-Pd-Mn have been motivated by the early discovery of
the stable icosahedral $i$-Al-Pd-Mn phase. In spite of the above
mentioned variety of magnetic properties found in these QCs, they
seem to exhibit the common feature of only a small fraction of the
manganese ions carrying a magnetic
moment\cite{QC:GavilanoRau,QC:Chernikov}. In some cases this
fraction even seems to decrease with decreasing
temperature\cite{QC:GavilanoRau}. Theoretical studies and
band-structure calculations aimed at explaining the phenomenon
that only a fraction of the Mn ions carries a magnetic
moment\cite{QC:GuyTramblyDeLaissardiere}, and to relate it to the
Hume-Rothery type pseudo gap in the electronic spectrum around the
Fermi level\cite{QC:HafnerKrajci}.

In comparison with the icosahedral Al-Pd-Mn alloys, the situation
is less clear for decagonal Al-Pd-Mn alloys. For the latter, a
larger number of sites has been predicted to carry a moment which,
however, is smaller than the Mn moment established in icosahedral
compounds\cite{QC:HafnerKrajci}. The present work reports the
results of measurements of the dc magnetic susceptibility and of
$^{27}$Al- and $^{55}$Mn-NMR studies of the decagonal QC
$d$-Al$_{69.8}$Pd$_{12.1}$Mn$_{18.1}$. From our results we infer
that the Mn-moments in this compound are distributed very
inhomogeneously. In the paramagnetic state and above 100K, only
about half of the Mn ions carry a magnetic moment. The data also
indicate that the fraction of Mn moments may even decrease upon
reducing the temperature below 100K. Finally we confirm the
previously reported spin-glass transition at $T_f
=12$K\cite{QC:ChernikovBeeli}.

\subsection{\label{sec:sample}The Sample}

A metastable icosahedral sample of Al-Pd-Mn was obtained by
spinning a melt of the stoichiometric composition of the decagonal
phase Al$_{69.8}$Pd$_{12.1}$Mn$_{18.1}$ onto a fast turning
(30m/s), water cooled Cu wheel. Annealing the resulting tapes with
the icosahedral phase for 100 hours at about 820$^{\circ}$C
transforms them into polycrystalline flakes of the stable
decagonal phase\cite{QC:BeeliStadelmann}. Decagonal Al-Pd-Mn has a
columnar structure arranged on a two-dimensional quasiperiodic
lattice. The translational period of length $d =
12.56\mathrm{\AA}$ is built up by two different types of layers: a
puckered layer P and a flat layer F\cite{QC:SteurerHaibach}. The
ratio of Al in P to Al in F is approximately 2:1; for Mn this
ratio is about 1:8\cite{QC:BeeliHoriuchi}.

\begin{figure}[ht]
\includegraphics[width=0.85\linewidth]{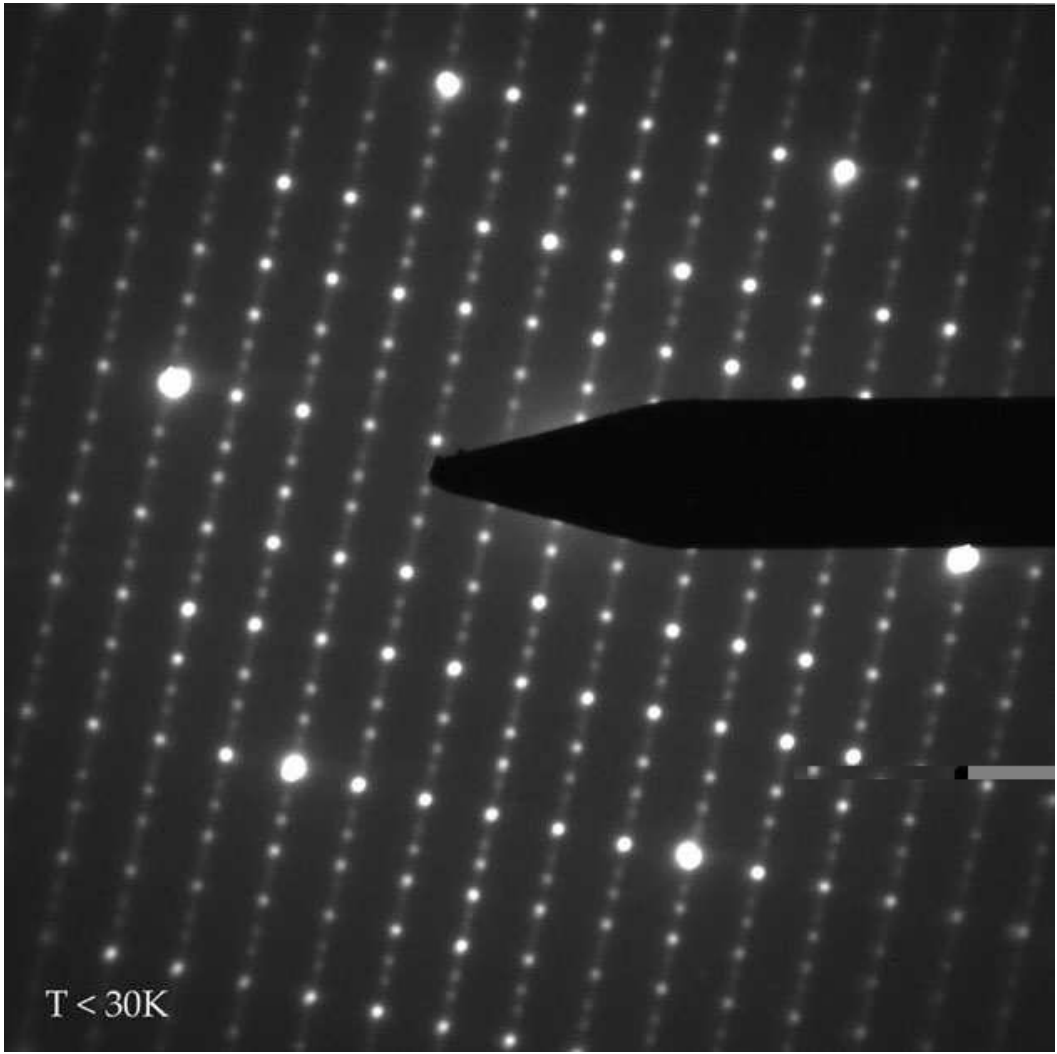}
\caption{\label{fig:SAED} Selected area electron diffraction
(SAED) picture of $d$-Al-Pd-Mn at 30K }
\end{figure}

High resolution transmission electron microscopy confirmed that
our sample consists of polycrystalline
$d$-Al$_{69.8}$Pd$_{12.1}$Mn$_{18.1}$ with no linear phason
strains. Selected area electron diffraction from room temperature
down to below 30K proved the high perfection of the decagonal
symmetry in the planes and the periodicity along the $c$-axis(see
figure \ref{fig:SAED}). Scanning electron microscopy sets the
upper limit of the admixture of a second phase, i.e.,
Al$_{11}$(Mn,Pd)$_{4}$ on the surface of the back side of the
tapes, to about 1\%.

\subsection{Experimental Results and their Analysis}
\subsubsection{\label{sec:susc}Magnetic Susceptibility}

\begin{figure}[ht]
\includegraphics[width=0.85\linewidth]{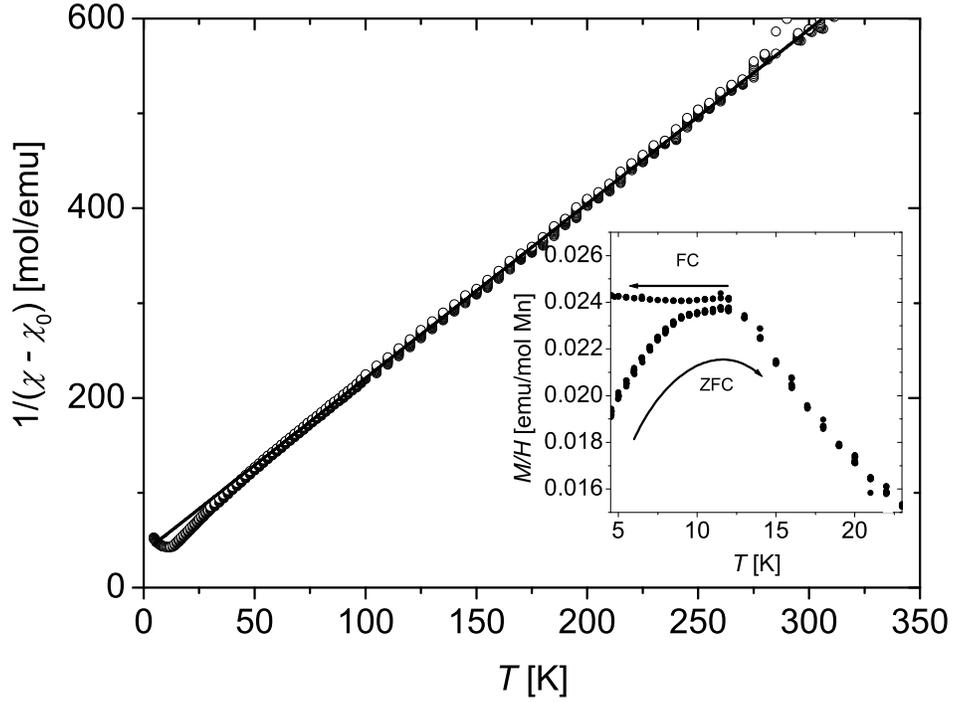}
\caption{\label{fig:suscept_500G} The inverse dc susceptibility
$\chi(T)$ after subtracting a constant negative offset per mol of
Mn in an applied field of 500G. The solid line represents the
high-temperature Curie-Weiss fit. The inset displays the dc
susceptibility below 24K. Below $T_f=12$K the zero field cooled
data deviates from the temperature independent field cooled data.}
\end{figure}

We measured the dc-susceptibility $\chi(T)$ of
$d$-Al$_{69.8}$Pd$_{12.1}$Mn$_{18.1}$ with a dc-SQUID magnetometer
between 340K and 2K, and at different magnetic fields between 50G
and 5.5T. Figure \ref{fig:suscept_500G} shows the inverse
susceptibility $1/(\chi(T)-\chi_0)$ measured at 500G between 5K
and 350K. As may be seen in the inset of figure
\ref{fig:suscept_500G}, field-cooled (FC) and zero-field cooled
(ZFC) data differ below $T_f \approx 12$K. There is a well-defined
maximum for the ZFC $\chi(T)$ at $T_f$. This behaviour confirms a
spin-glass type freezing of the Mn-moments, previously claimed
from the analysis of ac-susceptibility and specific-heat results
measured on the same sample\cite{QC:ChernikovBeeli}. Above 120K,
$\chi(T)$ may be approximated by $\chi(T)=C/(T-\Theta)+\chi_0$
with

\begin{equation}
C = N_A c p^2 \frac{\mu_B^2}{3k_B},
\end{equation}

where $N_A$ is Avogadro's number, $c$ the concentration of
magnetic ions, and $p$ their effective moment
$p=g(JLS)\sqrt{J(J+1)}$. The susceptibility data in the
temperature range between 120K and 350K are best approximated with
the parameters $C = 0.544$[emu/mol], $\Theta = -20$K, and $\chi_0
= -2.45\cdot 10^{-8}$[emu/g].

The negative paramagnetic Curie temperature $\Theta$ indicates a
predominant antiferromagnetic coupling between the Mn $d$-moments.
The Curie constant $C$ implies that $p\sqrt{c} = 2\pm 0.1 \mu_B$
per Mn-atom. This is not compatible with the assumption that all
Mn ions adopt a well-defined and identical ionic configuration. If
we assume, for example, localized Mn moments, our result for
$\chi(T)$ imply a concentration of only 16.5$\pm$0.5\% of
Mn$^{3+}$, or 12.5$\pm$0.5\% of Mn$^{2+}$. As we shall see in
section \ref{sec:spec}, however, about half of the Mn ions carry a
magnetic moment, and the distribution of the magnetic sites in
$d$-Al-Pd-Mn is very inhomogeneous. Distinct deviations of
$\chi(T)$ from the Curie-Weiss behaviour below 60K(see main frame
of figure \ref{fig:suscept_500G}) signal precursor effects of the
spin-glass freezing of the Mn moments.

\begin{figure}[ht]
\includegraphics[width=0.85\linewidth]{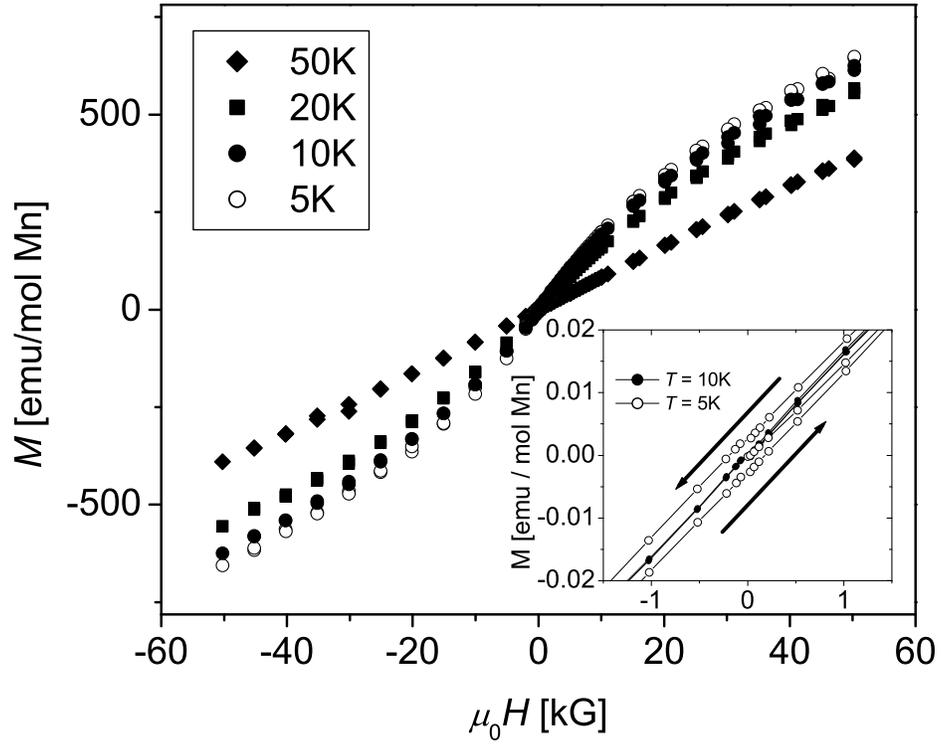}
\caption{\label{fig:magnetization}  Magnetization $M(H)$ at
various temperatures. Only the data at 5K shows a small
hysteresis, visible in the $M(H)$-curves in the inset.}
\end{figure}

Below 50K the susceptibility is field-dependent in fields below
5T. Figure \ref{fig:magnetization} displays the magnetization
$M(H)$ measured at various temperatures. As expected from the
$\chi(T)$ data there is a hysteresis at 5K, due to "memory
effects" in the spin-glass phase. There is no visible hysteresis
in the magnetization loop at 10K, suggesting that the relaxation
of the magnetization at that temperature is faster than the time
scale of the stepwise variation of the applied magnetic field in
our experiments.

\subsubsection{\label{sec:spec}NMR spectra}

\begin{figure}[ht]
\includegraphics[width=0.85\linewidth]{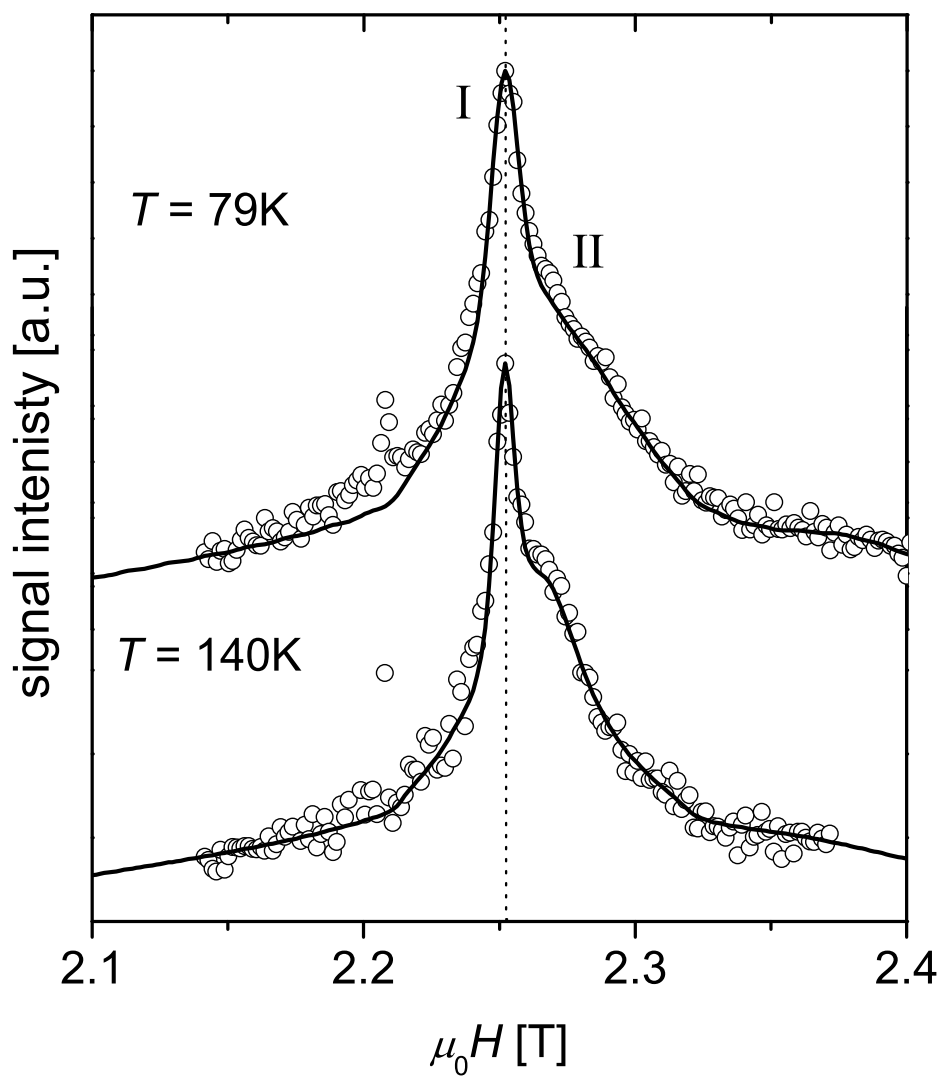}
\caption{\label{fig:tempdep-spectra_25MHz} $^{27}$Al NMR spectra
of $d$-AlPdMn at 24.97MHz with $\tau = 30\mu$s. The vertical
dotted line indicates the position of the $^{27}$Al line in an
aquaeous solution of AlClO$_3$. The solid lines result from
computer simulations of the spectra(see text). The narrow signal
at 2.207T arises from the $^{63}$Cu nuclei of the measurement
coil.}
\end{figure}

We recorded $^{27}$Al and $^{55}$Mn NMR spectra between room
temperature and 8K and at the frequencies of 21, 25, 30, 58, and
70MHz, using standard $\pi/2$-$\tau$-$\pi$ spin-echo sequences.
Figure \ref{fig:tempdep-spectra_25MHz} shows two spectra recorded
at 79 and 140K, respectively, with an excitation frequency of
24.97MHz. The NMR signal is distributed over a wide range of
resonant fields from 2.1 to 2.4T. The prominent peaks near 2.25T
represent the central $^{27}$Al Zeeman transition
($1/2\leftrightarrow -1/2$). The broad distribution of quadrupolar
wings is a generic feature of QCs\cite{QC:GavilanoAmbrosini}. The
vertical dotted line in Figure \ref{fig:tempdep-spectra_25MHz}
indicates the position of the resonance of $^{27}$Al nuclei in an
aquaeous AlClO$_3$ solution at room temperature and 24.97MHz,
which is used as a reference signal. The solid lines represent the
results of a computer simulation of the spectra which is described
below.

The $^{27}$Al central transition is very broad, almost a factor of
10 broader than expected from considering the second order
quadrupolar perturbation of the Zeeman line and than the analogous
signals observed for the case of the non-magnetic
Al-Pd-Re-compounds\cite{QC:GavilanoAmbrosini}. The line-width
increases with decreasing temperature and, as we shall see below,
the peculiar shape of the central line does not arise from
anisotropies in the Knight-shift or quadrupolar effects. We argue
that this signal consists of two partially resolved contributions
representing the central transitions of Al nuclei located in
either of two distinctly different local environments. One of the
contributions, denoted as line I in figure
\ref{fig:tempdep-spectra_25MHz}, has very small or zero
line-shift, while the other, denoted as line II in figure
\ref{fig:tempdep-spectra_25MHz}, exhibits a relatively large and
$T$-dependent negative shift.

\begin{figure}[ht]
\includegraphics[width=0.85\linewidth]{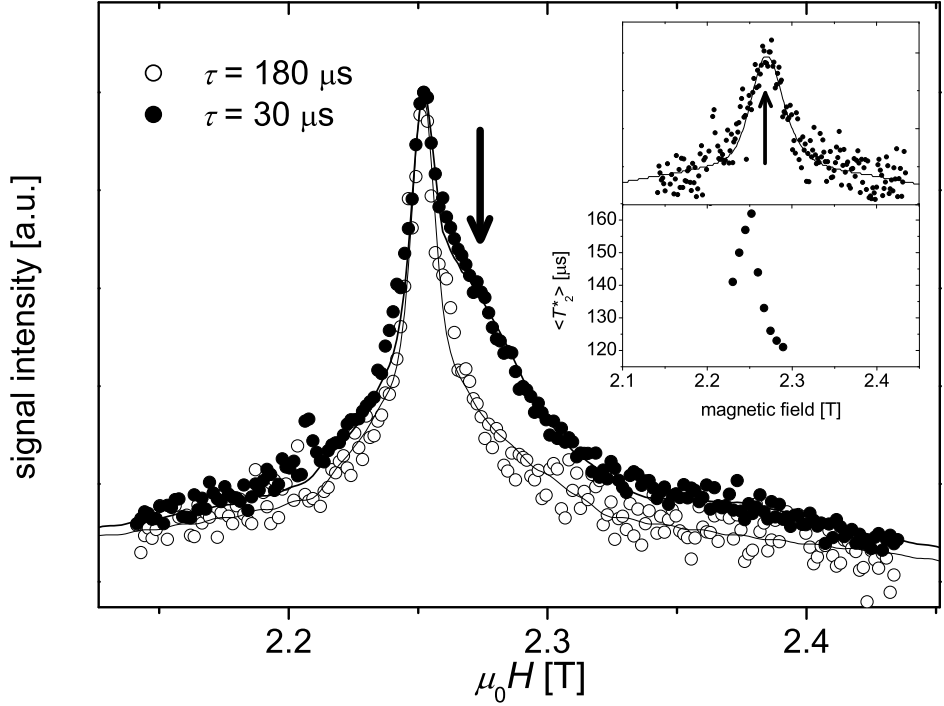}
\caption{\label{fig:taudep-spectra_25MHz} $^{27}$Al and $^{55}$Mn
NMR spectra of $d$-AlPdMn at 24.97MHz and 96K with $\tau = 30\mu$s
and $\tau = 180\mu$s. The arrow indicates the center of line II
which is also indicated by an arrow in the upper inset. The upper
inset displays the difference between the spectra at 180$\mu$s and
30$\mu$s, after taking $T_2^*$-effects into account. The solid
line in the inset is the contribution of Line II to fitting the
set of data represented by the full circles in the main frame. The
lower inset presents the average spin-spin relaxation rate
$<T_2^*>$ over the central part of the spectra. The excess signal
at 2.207T again arises from the $^{63}$Cu nuclei of the
measurement coil. For the definition of $T_2^*$ see text.}
\end{figure}

In order to confirm this conjecture we have looked for and found
appreciable differences in the spin-spin relaxation for the
$^{27}$Al nuclei that were allocated to different environments.
This was achieved by changing $\tau$ in the $\pi/2-\tau-\pi$ echo
sequence. Figure \ref{fig:taudep-spectra_25MHz} displays two
spectra recorded at 24.97MHz and 96K with two different values for
$\tau$. The spectrum with $\tau = 30\mu$s shows a distinct
shoulder on the high-field side while in the spectrum with
$\tau=180\mu$s this feature is clearly absent. In the upper inset
of figure \ref{fig:taudep-spectra_25MHz} we display the difference
of the two spectra after taking into account $T_2^{*}$-effects
($1/T_2^{*}=1/T_1+1/T_2\approx 1/T_2$). These effects affect both,
line I and line II, which experience a suppression of signal
intensity due to transverse relaxation processes. The loss of
signal due to these processes has been taken into account by
weighing factors obtained from a fit of $A \cdot
e^{\frac{-2\tau+d_2}{T^{*}_{2,\mathrm{I}}}}+ B \cdot
e^{\frac{-2\tau+d_2}{T^{*}_{2,\mathrm{II}}}}$ to the echo
intensity as a function of $\tau$, with $d_2$ as the duration of
the $\pi$-pulse. The thus obtained difference of the two spectra
then reveals a clear manifestation of line II. The solid lines in
the inset as well as in the main frame display the results of
computer simulations, briefly discussed below. We conclude that
the agreement between the experimental data and the calculated
curves provides strong evidence for the existence of two different
types of environments for the Al nuclei, giving rise to signals I
and II, respectively.

Support for our claims is obtained from the results of our
computer simulations of NMR spectra measured in a wide range of
$T$-, $H$- and $\tau$-values. The NMR frequency of a nucleus
depends on the local magnetic field and on the local electric
field gradient at the position of the nucleus, such that

\begin{equation}
\nu = \nu_{\mathrm{L}} - \nu_{\mathrm{Q}} (m-\frac{1}{2}) \frac{3
\mathrm{cos}^2 \theta -1}{2},
\end{equation}

where $\nu_{\mathrm{L}}$ and $\nu_{\mathrm{Q}}$ denote the Larmor
and the quadrupolar frequency, respectively, $m$ is the
$z$-component of the upper nuclear spin state of the transition,
and $\theta$ is the angle between the principal axis of the field
gradient with respect to the local magnetic field at the position
of the nucleus. We assumed that the Larmor frequencies
$\nu_{\mathrm{L}}$ obey a Gaussian distribution whose width $w$ is
caused by a ($T$-dependent) distribution of shifts, centered
around $\nu_0$. Our previous attempts for fitting the -- much
simpler -- $^{27}$Al-spectra of non-magnetic icosahedral
quasicrystals of the Al-Pd-Re-family were successful when choosing
the quadrupolar frequency $\nu_{\mathrm{Q}}$ to be uniformly
distributed between some hundred kilohertz and a few megahertz. In
the simulated $^{27}$Al-spectra of the present work, this
distribution of $\nu_{\mathrm{Q}}$ ranges from 200kHz to 2MHz. The
experimental spectra were successfully fitted by assuming two
different sets of environments, characterized by two Gaussian
distributions with different widths, $w_{\mathrm{I}}$ and
$w_{\mathrm{II}}$, spread around two different central frequencies
$\nu_{\mathrm{0,I}}$ and $\nu_{\mathrm{0,II}}$. The distribution
of $\nu_{\mathrm{Q}}$ cited above was assumed to be the same for
both environments. Figure \ref{fig:simspec70MHz} shows the
calculated three contributions to the total spectrum, i.e., line I
and line II of $^{27}$Al in the two different environments, and
the signal originating from nuclei of non-magnetic Mn
ions(discussed below). In order to compare the simulated spectra
with the experimental data, the frequencies were translated into
the corresponding applied fields. The distribution of angular
orientations of the local quadrupolar axes was assumed to be
random, because we used a polycrystalline sample.

\subsubsection{\label{sec:shifts}NMR line shifts and widths}

\begin{figure}[ht]
\includegraphics[width=0.85\linewidth]{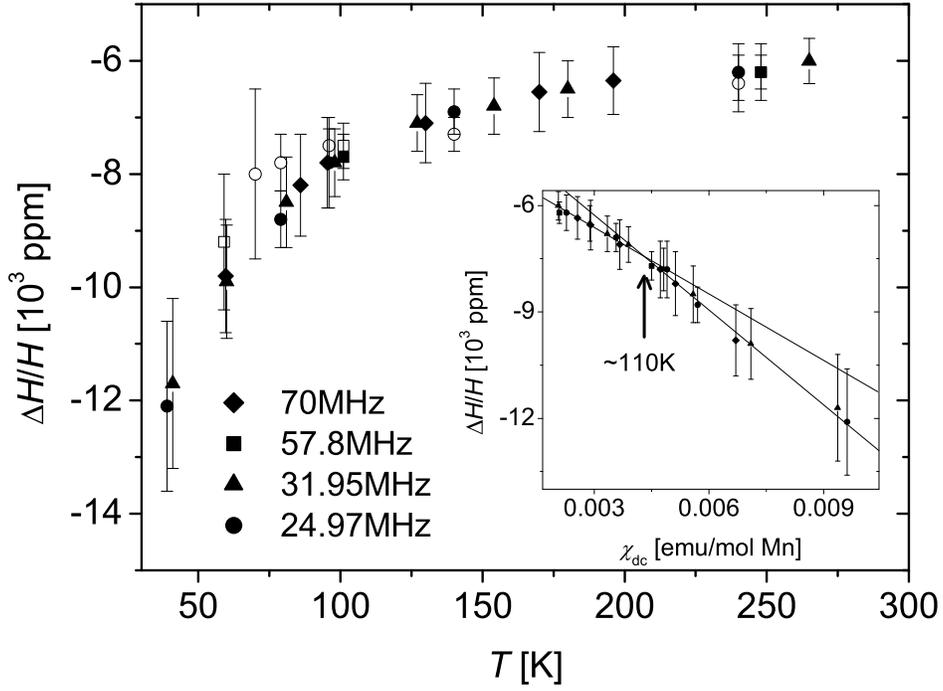}
\caption{\label{fig:KS} Temperature-dependence of the relative
$^{27}$Al line-shifts $\Delta H /H$ of line II in $d$-AlPdMn at
various fields. The open symbols are values obtained from the
difference of the lines recorded with $\tau=30\mu$s and
$\tau=180\mu$s, respectively. The closed symbols are values
obtained from fits to the spectra. The inset shows the line-shifts
plotted versus the dc-susceptibility. There is a slight change of
slope at approximately 110K, indicated by the two lines and the
arrow.}
\end{figure}

We determined the shift $\Delta H_{\mathrm{II}}$ of line II by
means of both, computer simulations of signals obtained with
$\tau=30\mu$s, and also by subtracting the spectrum monitored with
a long $\tau$ from that recorded with a short $\tau$, both at the
same temperature, of course. Figure \ref{fig:KS} shows the
relative line-shift $K_{\mathrm{II}} := \Delta H_{\mathrm{II}} /
H$ of line II as a function of temperature. The open symbols
indicate the values obtained by subtraction, the closed symbols
correspond to values obtained by computer simulations. The
absolute line shifts $\Delta H$ scale linearly with field and can
be expressed by

\begin{equation}
   K := \frac{\Delta H}{H} = K_{\mathrm{ce}} +
   K_{\mathrm{mag}}(T),
\end{equation}

where $K_{\mathrm{ce}}$ is a temperature-independent contribution,
which in our case is much larger in magnitude than those found in
non-magnetic
quasicrystals\cite{QC:GavilanoRau,QC:GavilanoAmbrosini,QC:ApihKlanjsek}.
Temperature-independent NMR line-shifts can be caused by a variety
of processes, however, $K_{\mathrm{ce}}$ of line II is too large
to be attributed to typical chemical shifts or to quadrupolar
effects. Therefore one is tempted to associate $K_{ce}$ with the
paramagnetism of the conduction electrons. In common metals this
contribution, the Knight shift, is positive, but, in our case
$K_{ce}$ is negative. A negative and temperature-independent line
shift cannot be caused by conduction electrons originating from
the Al $s$ and $p$ shells but may be due to itinerant electrons
originating from the Mn $d$-shell\cite{NMR:Clogston,book:Carter}.
The Pd $d$-band is full and therefore these states are not
expected to contribute significantly to the resonance shift. Such
an interpretation is compatible with recent experimental results
concerning the electronic structure of Al-Pd-Mn quasicrystals
\cite{QC:ErbudakRoessner}. Nevertheless, the origin of the
negative Knight-shifts, also found in other quasicrystals, is
still an open question. For line I the relative shift $K$ is also
negative but less than 100ppm. The existence of two distinct lines
with different shifts therefore strongly suggests that the
conduction electron density at the $^{27}$Al nuclei is not
uniform. In some regions, which contribute to the intensity of
line II, there is a substantial conduction electron density and,
as we argue below, the Mn ions in that region are magnetic. In
other regions, which contribute to the intensity of line I, the
conduction-electron density is reduced and the Mn ions are
non-magnetic.

$K_{\mathrm{mag}}(T)$, only clearly observed for line II, exhibits
a Curie-Weiss type behaviour in the whole temperature range
covered by the data shown in figure \ref{fig:KS}. The paramagnetic
Curie-temperature $\theta=-6$K is in fair agreement with the value
obtained from fits to the susceptibility. The inset of figure
\ref{fig:KS} displays the line-shift data plotted versus the
dc-susceptibility measured in different fields. It confirms that
$K_{\mathrm{mag}}(T)$ may be written as

\begin{equation}
   K_{\mathrm{mag}}(T) = A \cdot \chi(T).
\end{equation}

The proportionality constant $A$ is a measure of the average
coupling strength of the $^{27}$Al nuclear spins to the Mn
magnetic moments. For metals, $A$ is usually written as

\begin{equation}
\label{eq:A}
   A = \frac{1}{\mu_B N}
   H_{\mathrm{eff}},
\end{equation}

with $H_{\mathrm{eff}}$ as the hyperfine field from each aligned
Bohr magneton. $N$ is Avogadro's number\cite{book:Carter}. In our
case, where only a fraction $c$ of the Mn ions is magnetic and
because the molar susceptibility is calculated per total Mn
content, equation (\ref{eq:A}) has to be changed to

\begin{equation}
   A = \frac{1}{\mu_B c N}
   H_{\mathrm{eff}}.
\end{equation}

In spite of the large error bars, our data suggest that there are
two temperature regimes with different values of $A$, below and
above approximately 110K. This change may simply reflect a
reduction of the fraction of magnetic Mn moments $c$. This
interpretation gains support from the $^{55}$Mn NMR spectra, where
we observe that the intensity of the \textit{non-magnetic} Mn line
is 35\% and 50\% of the total expected Mn intensity above and
below 100K, respectively. However, other processes leading to a
change in the hyperfine coupling cannot be ruled out.

\begin{figure}[ht]
\includegraphics[width=0.85\linewidth]{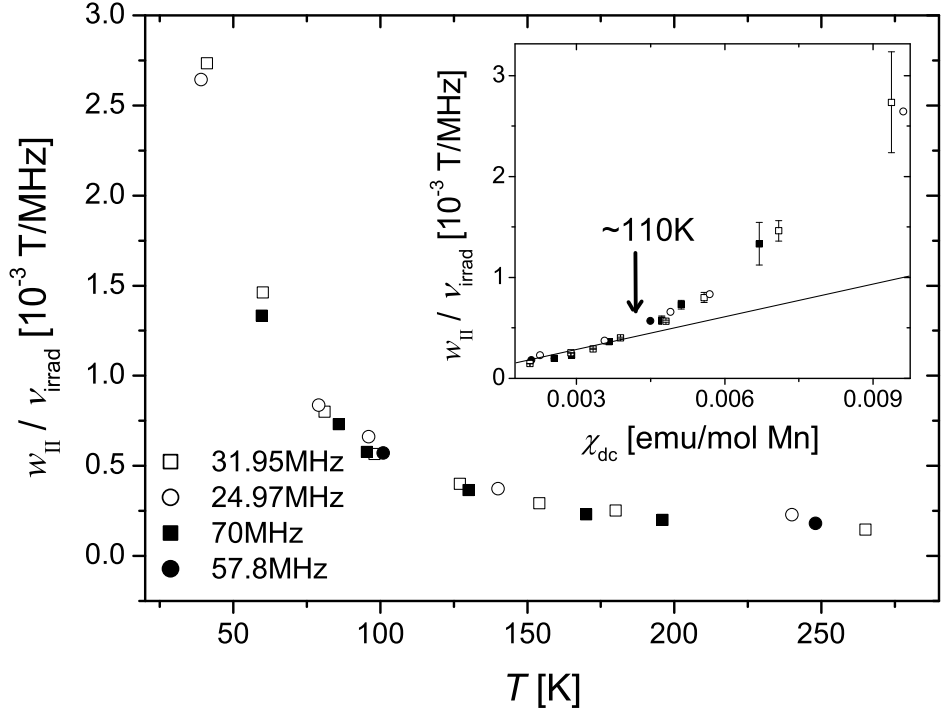}
\caption{\label{fig:widthII} Temperature-dependence of the width
of line II divided by the irradiation frequency
$\nu_{\mathrm{irrad}}$. The inset shows the same ratio plotted
versus the dc-susceptibility. There is a pronounced change of
slope at approximately 110K, indicated by the arrow. The solid
line fits the data at elevated temperatures.}
\end{figure}

The widths of both lines increase with decreasing temperature and
therefore give a clear indication that they are both of magnetic
origin. Also line I is much broader than the central $^{27}$Al
line in other Al-Pd-Mn compounds\cite{QC:GavilanoRau}, suggesting
that the regions of the two environments mentioned above are
finely dispersed in the bulk of the sample. At this point we focus
our attention on line II, where the influence of the Mn moments is
particularly pronounced. In figure \ref{fig:widthII} the
temperature dependence of the ratio
$w_{\mathrm{II}}/\nu_{\mathrm{irrad}}$, where $w_{\mathrm{II}}$ is
again the width of line II and $\nu_{\mathrm{irrad}}$ the
corresponding irradiation frequency, is plotted for four different
irradiation frequencies, i.e., four different average applied
fields. Since all the data fall onto a single curve, it follows
that the line-width scales with the irradiation frequency, or,
equivalently, with the applied magnetic field and, therefore, must
be of magnetic origin. The enhancement with decreasing temperature
occurs with increasingly negative slope. The inset of figure
\ref{fig:widthII} displays the same data plotted versus the
dc-susceptibility measured at similar fields. The width of line II
originates from a distribution of shifts of the $^{27}$Al nuclei
around an average shift. The slope of that curve therefore is a
measure of the width of the distribution of the site dependent
coupling strengths $A(\mathrm{\bf r})$ between the Al nuclear
spins of line II and the Mn moments, with respect to the total
magnetic susceptibility, which is mainly due to localized Mn
moments. It shows a pronounced change at approximately 110K. Such
a slope change could be induced by a reduction of the number of
magnetic Mn ions, leading to a reduced increase in the
susceptibility.

\begin{figure}[ht]
\includegraphics[width=0.85\linewidth]{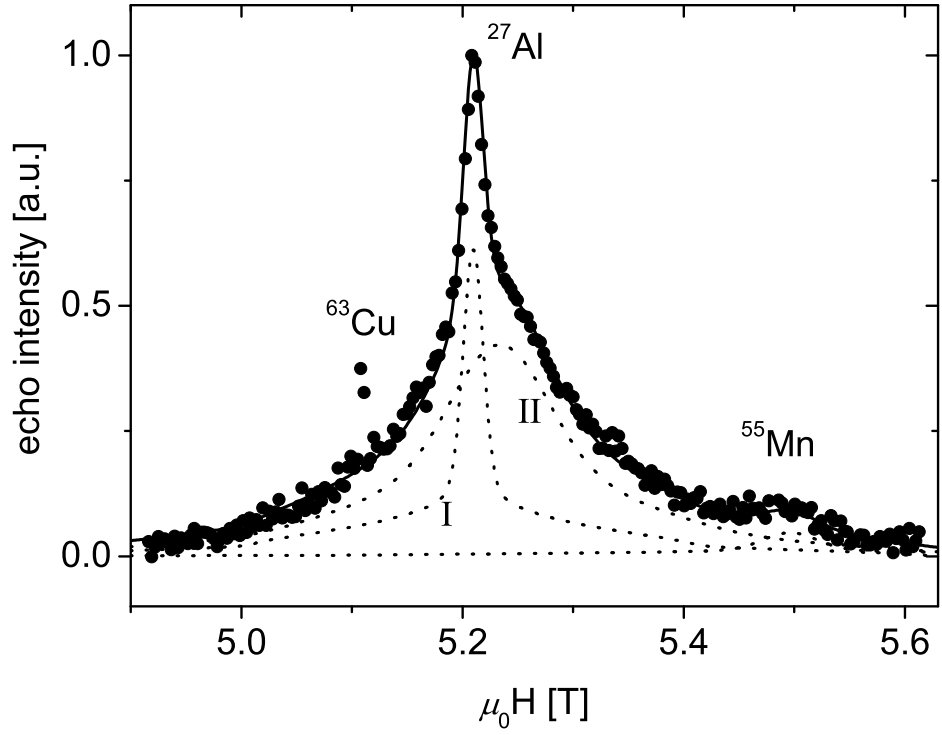}
\caption{\label{fig:simspec70MHz} $^{27}$Al and $^{55}$Mn NMR
spectrum of $d$-AlPdMn at 57.8MHz and 101K. The solid line
represents a fit to the data, the dotted lines indicate the
individual contributions of line I, line II and the Mn-signal.}
\end{figure}

The NMR spectra recorded at a chosen frequency reveal some
additional intensity at a field where one expects the $^{55}$Mn
signal at zero shift. This is most clearly seen near 5.5T in
figure \ref{fig:simspec70MHz}. Since the hyperfine field coupling
of the Mn moment to the nucleus of the same ion is through the
core-polarization, which is of the order of
-100kG/$\mu_{\mathrm{B}}$ for Mn\cite{QC:WarrenChen}, even a small
moment on a Mn ion would shift its resonance frequency by several
MHz. Therefore this signal intensity in our spectra must originate
from non-magnetic Mn atoms (see figure \ref{fig:simspec70MHz}).
From the spectra recorded at 70MHz, where the Mn-signal is clearly
resolved, we conclude that the fraction of Mn-atoms carrying no
magnetic moment is of the order of 45$\pm$15\%. We note that this
value is not far from the ratio of the intensities of line I and
II which was found to be $0.4\pm0.1$. Thus we may again conclude
that the Al lines I and II arise from Al nuclei in the
neighborhood of nonmagnetic and magnetic Mn ions, respectively.
Considering the average moment of $p = 2\pm 0.1 \mu_B$, deduced
from the dc molar susceptibility data with respect to the total Mn
content, we estimate that the average effective moment per
magnetic Mn ion is $p = 3\mu_B$, if we assume that 45\% of the Mn
ions actually are magnetic. This number is not compatible with any
of the possible ionization states of Mn, which would lead to
$p=4\mu_B$ for Mn$^{4+}$, $p=5\mu_B$ for Mn$^{3+}$, and
$p=5.9\mu_B$ for Mn$^{2+}$.

\subsubsection{\label{sec:relaxation}NMR Spin-lattice relaxation rate}

We measured the spin-lattice relaxation rate $T_1^{-1}$ of line I
from room temperature down to 15K in three different applied
fields. The values of $T_1$ were extracted from fits to the
nuclear magnetization recovery $m(t)$, which was first destroyed
using a long comb of rf-pulses, followed by a variable delay $t$
and a spin-echo sequence $\pi/2-\tau-\pi$ with a long
$\tau=180\mu$s. In this manner it is possible to identify
exclusively the spin-lattice relaxation rate related to line I.
Using shorter values of $\tau$ in these measurements leads to
changes in $m(t)$ which, in our view, results in less reliable
values of $T_1$. According to standard NMR theory, the signal
intensity of the central line of a spin $I=5/2$ nucleus relaxes
such that\cite{NMR:Narath}:

\begin{equation}
\label{eq:si}
1-m(t)/m(\infty) = 0.4762 e^{-15t/T_1} + 0.2667 e^{-6t/T_1} +
0.2571 e^{-t/T_1}.
\end{equation}

Even for line I alone, there is no distinct single Al-site, but
rather a variety of sites leading to a distribution of $T_1$'s.
Since the details of the distribution of $T_1$'s are not known, we
simply model $m(t)$ by replacing the exponentials by stretched
exponential functions\footnote{We found good agreement between
equation (\ref{eq:stretched}) and magnetization recoveries
obtained by computer simulations of a collection of ensembles with
different $T_1$.}:

\begin{equation}
\label{eq:stretched}
1-m(t)/m(\infty) = 0.4762
e^{-15(t/T_1)^\beta} + 0.2667 e^{-6(t/T_1)^\beta} + 0.2571
e^{-(t/T_1)^\beta}.
\end{equation}

\begin{figure}[ht]
\includegraphics[width=0.85\linewidth]{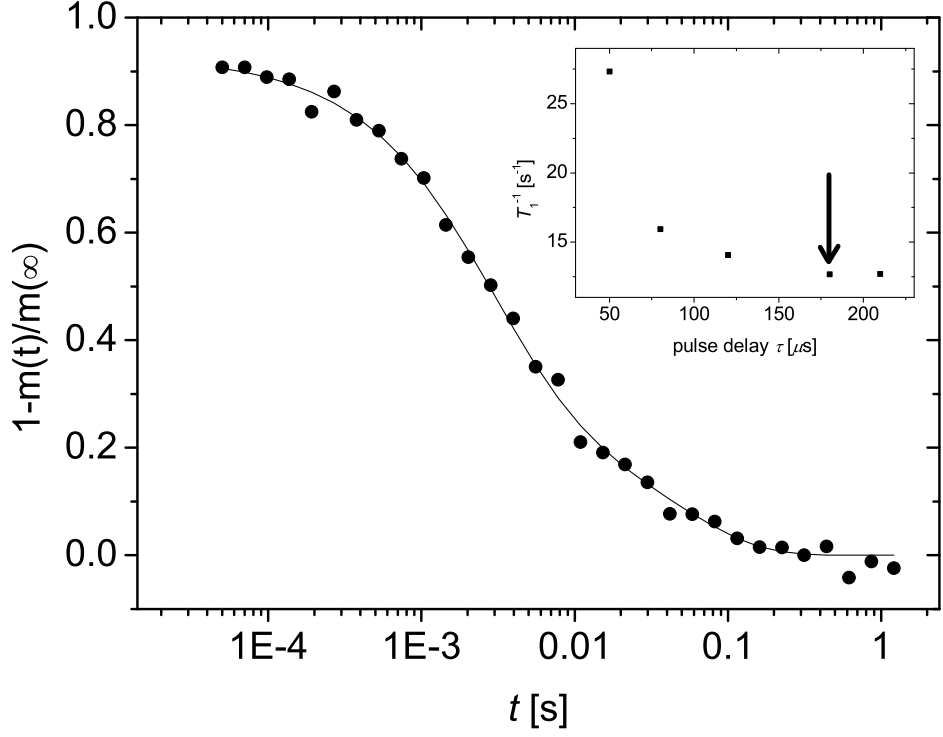}
\caption{\label{fig:mag_recov} Magnetization recovery curve
related with line I at 57.8MHz and 24.5K. The solid line
represents a fit to equation (\ref{eq:stretched}). The inset
displays how $T_1^{-1}(68\mathrm{K})$ saturates with increasing
$\tau$. The arrow marks $\tau=180\mu$s, the pulse delay that was
used in our measurements.}
\end{figure}

Figure \ref{fig:mag_recov} displays an example of a magnetization
recovery at a relatively low temperature and the corresponding
best fit to the data using equation (\ref{eq:stretched}). With an
increasing pulse separation $\tau$ in the echo sequence the
exponent $\beta$ converges to an approximate value of 0.9.
Therefore, in the subsequent analysis we set $\beta=0.9$. As may
be seen in the inset of figure \ref{fig:mag_recov}, also the $T_1$
values turned out to vary with $\tau$ but to reach a constant
value for $\tau \geqslant 180\mu$s. Therefore, we fixed $\tau$ to
$180\mu$s for all the measurements of $T_1$ discussed below.

\begin{figure}[ht]
\includegraphics[width=0.85\linewidth]{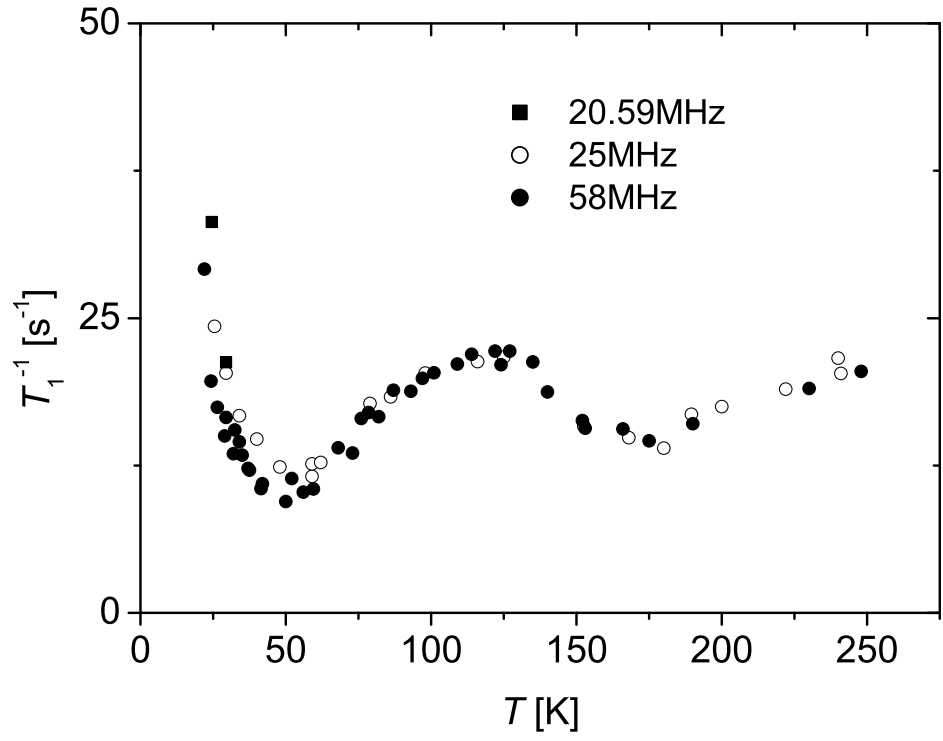}
\caption{\label{fig:T1} Spin-lattice relaxation rate $T_1^{-1}(T)$
for line I between 25 and 250K.}
\end{figure}

Figure \ref{fig:T1} displays $T_1^{-1}(T)$ obtained from fitting
the data obtained at three different fields of 5.2T, 2.25T, and
1.85T, with $\tau=180\mu$s to equation (\ref{eq:stretched}). Below
50K the spin-lattice relaxation rate increases with an
increasingly negative slope as $T$ approaches the spin-glass
freezing temperature $T_f$. This behaviour is typical for nuclear
spins which experience the influence of magnetic moments that
undergo a gradual slowing down of their fluctuations. In the
present case, the Mn moments, although predominantly affecting
line II, are also felt in the relaxation of the nuclear spins
contributing to line I, as the spin-glass transition is approached
upon cooling. At high temperatures, the spin-lattice relaxation
rate of non-magnetic nuclei due to paramagnetic centers is given
by\cite{NMR:Abragam}:

\begin{equation}
\label{eq:fluct} \frac{1}{T_1}  =  C \frac{\tau}{1
+\omega^2_I\tau^2}.
\end{equation}

$C$ is proportional to the hyperfine-field coupling between the Mn
moments and the Al nuclei, $\tau$ is the correlation time of the
Mn moments, and $\omega_I=2\pi\nu_{\mathrm{0,I}}$ is the Larmor
precession frequency of the nucleus under
investigation\footnote{Equation (\ref{eq:fluct}) is a
high-temperature approximation where the correlations among the
paramagnetic moments are neglected\cite{NMR:Abragam}.}. In the
fast-motion regime, at temperatures much higher than $T_f$,
$\omega_I^2\tau^2$ is much smaller than 1 and the relaxation rate
is temperature- and field-independent. As $T$ is reduced towards
$T_f$, the moment fluctuations slow down and cause an increase of
$T_1^{-1}$($\propto \tau$), as observed below 50K. The
$\tau$-driven increase can be understood by assuming that
$\omega^2\tau^2$ is still much less than 1, which is implicitly
confirmed by the almost $H$-independent increase of $T_1^{-1}(T)$
below 50K. Close to $T_f$, the growing linewidth usually prohibits
to extract reliable $T_1$ values. In our case this is true for
$T<20$K.

The $T_1^{-1}(T)$ plot in figure \ref{fig:T1} exhibits a broad
maximum centered around 120K. This maximum is very unusual and has
-- to our knowledge -- never been seen in $T_1^{-1}(T)$ of any
other quasicrystalline compound. The spin-lattice relaxation rate
usually contains three contributions

\begin{equation}
\label{eq:rates} T_1^{-1} = T_{1,\mathrm{ce}}^{-1} +
T_{1,\mathrm{mag}}^{-1} +T_{1,\mathrm{q}}^{-1}.
\end{equation}

\begin{figure}[ht]
\includegraphics[width=0.85\linewidth]{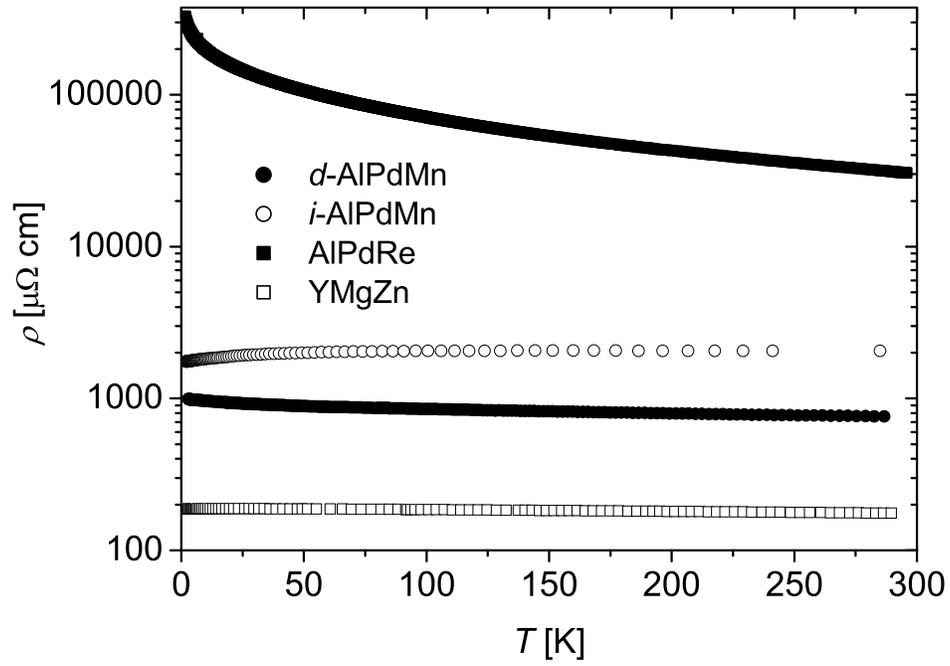}
\caption{ \label{fig:rho} Zero-field electrical resistivity
$\rho(T)$ of $d$-Al-Pd-Mn compared to $\rho(T)$ of $i$-Al-Pd-Mn
\cite{QC:Dolinsek}, $i$-Al-Pd-Re\cite{QC:BianchiBommeli}, and
Y-Mg-Zn\cite{QC:GiannoRE}.}
\end{figure}

The total relaxation is characterized by $T_{1,\mathrm{ce}}^{-1}$,
due to conduction electrons, $T_{1,\mathrm{mag}}^{-1}$ reflecting
the relaxation due to the paramagnetic centers, and
$T_{1,\mathrm{q}}^{-1}$ capturing the quadrupolar relaxation. In a
simple metal, $T_{1,\mathrm{ce}}^{-1}$ is known to vary linearly
with temperature. In quasicrystals the relaxation due to itinerant
electrons depends on the shape of the density of states at the
Fermi level and thus on the shape of the pseudogap. A power-law or
polynomial increase of the spin-lattice relaxation rate with
rising temperature is
expected\cite{QC:TangHill,QC:GavilanoAmbrosini}. It may be seen in
figure \ref{fig:T1} that above 175K the $T$-dependence of the
spin-lattice relaxation rate may well be dominated by conduction
electrons and they still contribute significantly to the total
$T_1^{-1}$ in the crossover regime around 120K. However
$T_{1,\mathrm{ce}}^{-1}$ cannot account for the maximum at 120K,
unless some anomaly in the electronic excitation spectrum causes
the feature at that temperature. Major changes in
$T_{1,\mathrm{ce}}$ around 120K are, however, not to be expected
because the monotonous variation of the electrical resistivity
with temperature, shown in figure \ref{fig:rho}, does not support
a corresponding feature in the electronic excitation spectrum that
would cause the anomaly in $T_1^{-1}$ around 120K.

Also any substantial contribution of $T_{1,\mathrm{q}}^{-1}$ can
be ruled out. The spin-lattice relaxation rate of Al nuclei is
larger by more than a factor of 10 in $d$-Al-Pd-Mn than it is in
$i$-Al-Pd-Re, where it is still dominated by
$T_{1,\mathrm{ce}}^{-1}$ and not
$T_{1_,\mathrm{q}}^{-1}$\cite{QC:TangHill}. Hence
$T_{1,\mathrm{q}}^{-1}$ in $d$-Al-Pd-Mn is expected to be very
small in comparison with other contributions, unless a major
change of structure at some temperature occurs in that compound.
Results of selected area electron diffraction (SAED) measurements
from room temperature down to below 30K indicate that no
structural changes occur in this temperature regime and therefore
it may be concluded that $T_{1,\mathrm{q}}^{-1}$ does not
contribute significantly to changes in $T_1^{-1}(T)$.

We are thus left with the second significant term in
(\ref{eq:rates}), i.e. $T_{1,\mathrm{mag}}^{-1}$, which describes
the magnetic interactions between the Al nuclear spins and the Mn
$d$-moments. In our case, with $\omega_{\mathrm{I}}\tau<<1$ in
equation equation (\ref{eq:fluct}), $T_1^{-1}$ is expected to be
almost $T-$ and $H-$independent. We note that the values of
$T_1^{-1}$ of other magnetic quasicrystals of the
Al-Cu-Fe-family\cite{QC:TangHill,QC:DolinsekKlajnsek} and
icosahedral samples of Al-Pd-Mn with lower Mn
concentration\cite{QC:GavilanoRau} are smaller than for the
presently investigated material by at least one order of magnitude
as well. This giant difference is mainly attributed to the larger
contribution of $T_{1,\mathrm{mag}}^{-1}$ to the total
spin-lattice relaxation rate in the present case. Based on these
arguments we suggest that the maximum in $T_1^{-1}(T)$ is related
to $T_{1,\mathrm{mag}}^{-1}(T)$ and reflects the previously
mentioned reduction of the concentration of magnetic Mn moments
below approximately 100K.

\begin{figure}[ht]
\includegraphics[width=0.85\linewidth]{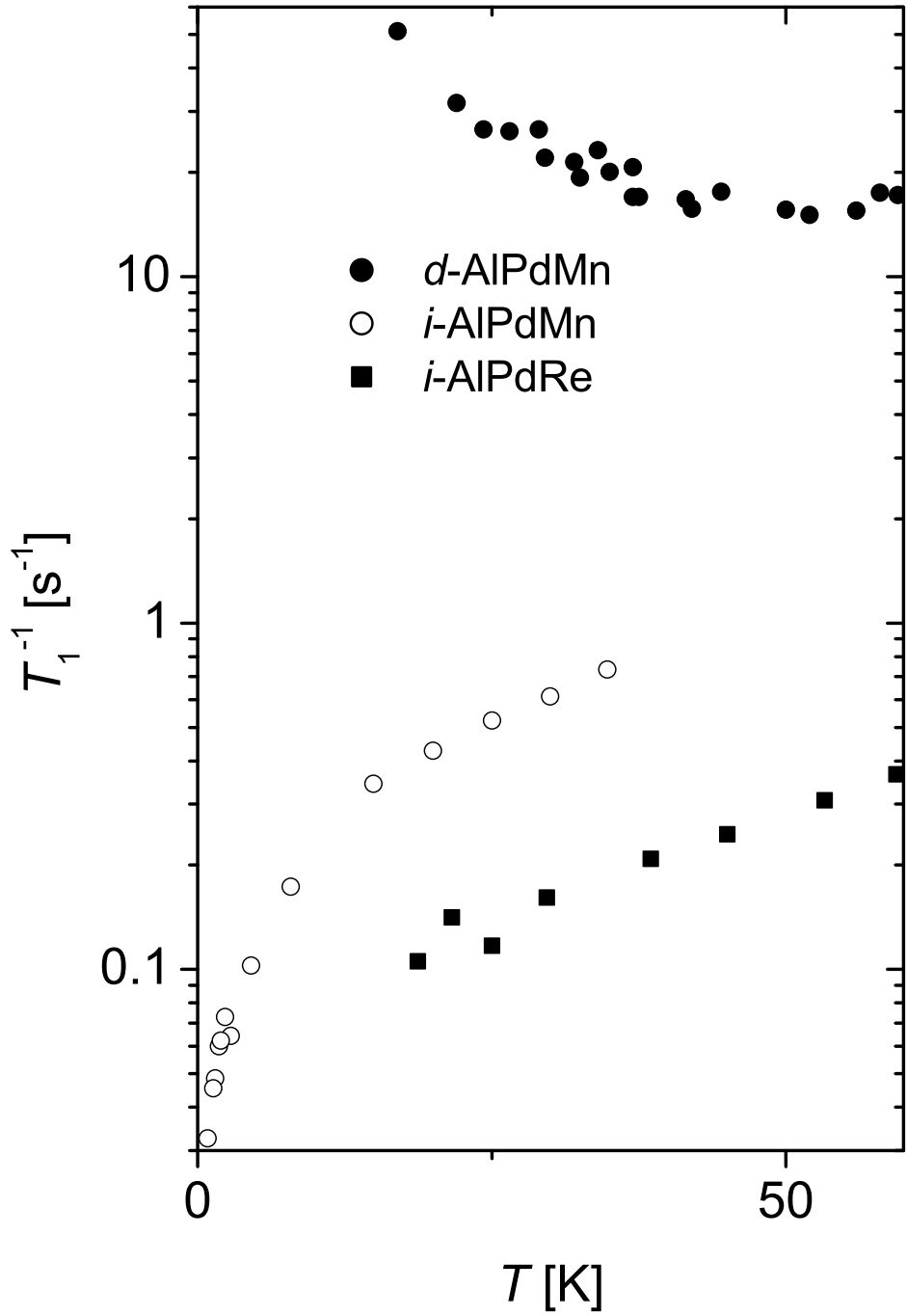}
\caption{\label{fig:relaxation_rates} Spin-lattice relaxation
rates $T_1^{-1}$ of $^{27}$Al in the non-magnetic icosahedral
quasicrystal $i$-AlPdRe\cite{QC:GavilanoAmbrosini}, in icosahedral
$i$-Al-Pd-Mn\cite{QC:GavilanoRau}, and in $d$-Al-Pd-Mn.}
\end{figure}

In figure \ref{fig:relaxation_rates} we compare the low
temperature $^{27}$Al spin-lattice ralaxtion rate of $d$-Al-Pd-Mn
with those of $^{27}$Al nuclei in a non-magnetic $i$-Al-Pd-Re
quasicrystal\cite{QC:GavilanoAmbrosini} and in
$i$-Al-Pd-Mn\cite{QC:GavilanoRau}. In the covered temperature
region the relaxation rates in both, the non-magnetic $i$-Al-Pd-Re
as well as the weakly magnetic $i$-Al-Pd-Mn, are smaller by more
than an order of magnitude and exhibit a distinctly different
temperature dependence.

\subsection{\label{sec:sum}Summary}

Our susceptibility data and our NMR spectra reveal that at
elevated temperatures more than half of the Mn ions in decagonal
Al$_{69.8}$Pd$_{12.1}$Mn$_{18.1}$ carry a small magnetic moment,
giving rise to an average Mn-moment of 2$\mu_{\mathrm{B}}$. These
moments experience an average antiferromagnetic interaction,
leading to a spin-glass freezing at $T_f\approx12$K. We presented
arguments that the number of these moments is gradually reduced
below 100K. A similar reduction of moments was observed at lower
temperatures in $i$-Al-Pd-Mn with a much smaller density of
magnetic Mn moments\cite{QC:GavilanoRau}.

The $^{27}$Al NMR spectra show two partially resolved lines. Line
II is clearly much more influenced by the Mn-magnetism. Similar
line-shapes have been found in very early experiments on Mn-rich
metastable Al-Mn quasicrystals\cite{QC:WarrenChen}, but they could
not be resolved into two lines at that time. We argue that the two
lines are due to Al on sites in environments favoring the
formation of a magnetic moment on the Mn ions (line II) and to Al
in an environment, where the Mn moments are quenched (line I). Our
results strongly suggest that simple Al-$p$-Mn-$d$ hybridization
alone may not adequately account for the phenomenon of
Mn-magnetism in QC's, but that a more subtle, conduction-electron
mediated long-range mechanism has to be considered
\cite{QC:GuyTramblyDeLaissardiere}.

We have not yet found an unequivocal interpretation for the broad
maximum of $T_1^{-1}$ at 120K. The same process that leads to that
maximum seems to lead to a reduction of the fraction of Mn-ions
that carry a magnetic moment, or, possibly an increased hyperfine
coupling below 110K. Neither a structural transition nor a
significant change in the electronic density of states is
indicated by SAED and transport measurements, respectively.

\subsection{\label{sec:acknow}Acknowledgement}

We thank Ren$\acute{\mathrm{e}}$ Monnier for fruitful discussions,
Roland Wessiken for his help with the temperature-dependent
selected area electron diffraction, and Matthias Weller for his
assistance in the measurement of the electrical conductivity.


\end{document}